\documentclass{notices}

\usepackage{amsfonts,amssymb,amsmath,amscd}

\usepackage{hologo}
\usepackage{paralist}

\usepackage{graphicx}

\usepackage{cleveref}

\title{
DONUT -- Creation, development, and opportunities of a database
}

\author{
    Barbara Giunti%
    \affil{
        Barbara Giunti is a postdoctoral researcher in the Kerber research group at Graz University of Technology. Her email address is bgiunti@tugraz.at.
    }
    \and
    J\=anis Lazovskis%
    \affil{
        J\=anis Lazovskis is a \textit{docents} at the Riga Technical University and a faculty member at RTU Riga Business School. His email is janis.lazovskis\_1@rtu.lv.
    }
    \and
    Bastian Rieck%
    \affil{
        Bastian Rieck is a Principal Investigator at Helmholtz Munich and a Faculty Member of the School of Computation, Information and Technology at Technical University of Munich. His e-mail is bastian.rieck@tum.de.
    }
}

\begin{document}

\maketitle

\section{Origin}

DONUT \cite{DONUT} is a database of papers about practical, real-world uses of Topological Data Analysis~(TDA). 
Its original seed was planted in a group chat formed during the HIM Spring School on Applied and Computational Algebraic Topology in April 2017.

\begin{figure}[h!]
\includegraphics[width=\columnwidth]{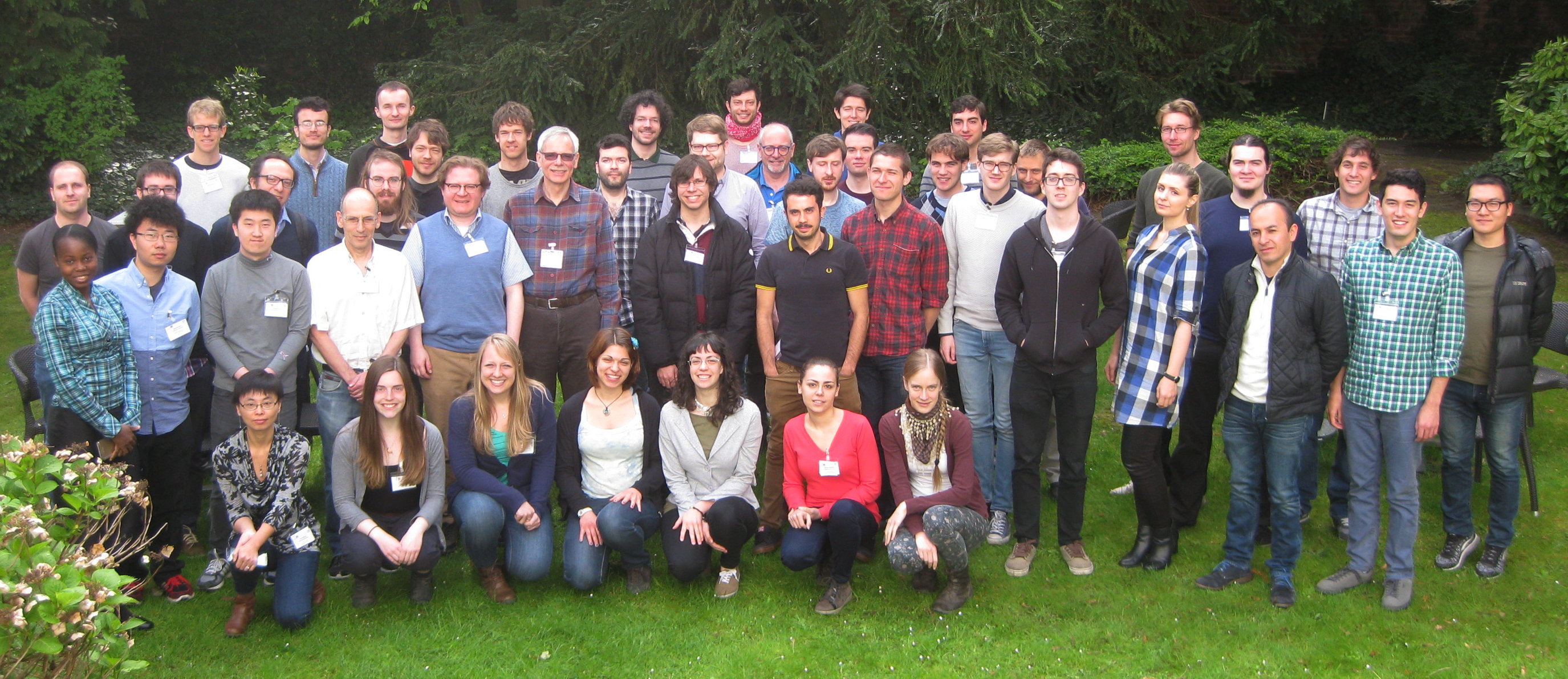}
\caption{Participants of the 2017 Spring School at the Hausdorff Research Institute for Mathematics, Bonn. Copyright HIM.}
\label{fig:him}
\end{figure}

In 2019, Barbara Giunti, at the time a PhD student at the University of Pavia, asked the group chat whether anyone had heard of applications of topology in a specific area. 
J\=anis Lazovskis, also a PhD student back then~(at the University of Chicago), had been collecting such papers during the spring school and later events, and shared a list of about 10 papers demonstrating TDA applications. 
The format of an online spreadsheet soon proved too restrictive, and in 2020, they moved to Zotero \cite{zotero_database}, an application specifically designed to handle bibliographic databases. 
The number of applications had increased by then to around 30, and J\=anis and Barbara started to feel the need for smart planning: what if this project will grow as we dream, with hundreds of entries and to be used by many? How can we make it searchable and versatile? 
They came up with a \textit{tags} and \textit{flavors} system compatible with the Zotero infrastructure and set to classify all the papers accordingly~(more information about the system can be found in \Cref{s_tags}). 
In the meantime, they started advertising the database, getting immediate positive feedback from the community. 
Among the backers, they got the great help of Prof.\ Mikael Vejdemo-Johansson, who provided more than one hundred papers from his personal database. 
Moreover, the year after, he covered the cost of the annual Zotero subscription since the volume of papers had by then exceeded the threshold of free storage (the fee has since been footed by Barbara). 

\begin{figure}[h!]
\includegraphics[width=\columnwidth]{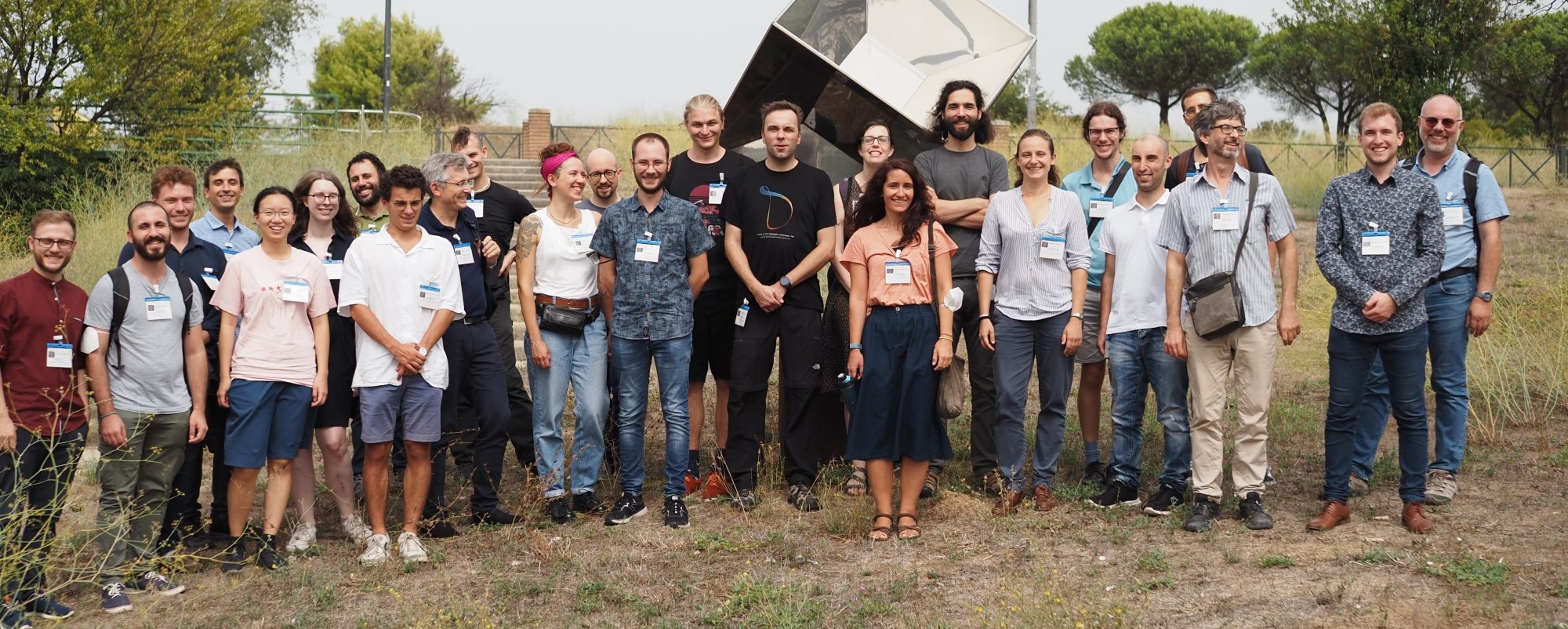}
\caption{Participants of the 2022 workshop Topology of Data in Rome. Copyright Uni Roma.}
\label{fig:roma}
\end{figure}

In 2022, the Zotero database had reached over 300 entries, all classified according to the tags and flavors system, and it began to show its limitations: one needed to be familiar with the app to successfully find the desired references.
Luckily, in Salzburg during the biannual TDA Austrian Meeting and then at the workshop ``Topology of Data in Rome", Barbara met Bastian Rieck, who fell in love with the project and revitalized it with his contribution: the web frontend search engine DONUT. This acronym stands for ``Database of Original \& Non-Theoretical Uses of Topology", and references one of the most basic shapes with non-trivial topology, the donut. 

\section{Motivation}

The original goal was to have a tool to retrieve all applications of topology in a specific domain, to provide an overview not only at the specific application level but also of all the areas of applications at a higher level. 
This tool is useful, for example, to promote the research field, showcasing its richness and power.
Having such a tool is particularly handy in preparing introductions of papers and theses and, crucially, when writing grant applications, to reference relevant uses of a particular method. 
It also serves as a way to attract more researchers to the field; computational topology being a highly intersectional and interdisciplinary field, it is open to new contributions from different domains.
DONUT is also useful to create or extend projects, for example, by applying TDA in novel domains or overcoming limitations of previous approaches. 

Another goal in creating this database was to organize existing knowledge, a burning necessity in an age of information overload. 
For this reason, the tags and flavors include not only the area of applications but also what mathematical tools are used, how the data are retrieved and pre-processed, and how novel the results are in the specific domain of application. 
Having the information in such a structured format not only helps the practitioners achieve their research goals, but also allows for literature and cross-sectional studies.
In the absence of a structured bibliographic format for reporting such details, DONUT serves the important purpose of providing an ever-evolving, dynamic taxonomy of the field.

\section{How the cataloging works}

\subsection{Admissibility criteria}

To be included in DONUT, an entry must use a TDA technique to analyze data. 
We therefore exclude applications to other areas of mathematics or computer science, or employing mathematical (even topological) tools that are not part of the TDA toolbox. 

The entries we index must be either published or available as a preprint on some preprints server (such as arXiv or bioRXiv).
We prefer open-access~(OA) publications and items with a DOI.
Preprints that are later published are replaced with their published version. 
If the later publication is not open access, a link to the public preprint is kept. Conference submissions are allowed only if published in proceedings; conference submissions that only consist of an abstract are not included.
%

\begin{figure}[h!]
\includegraphics[width=\columnwidth]{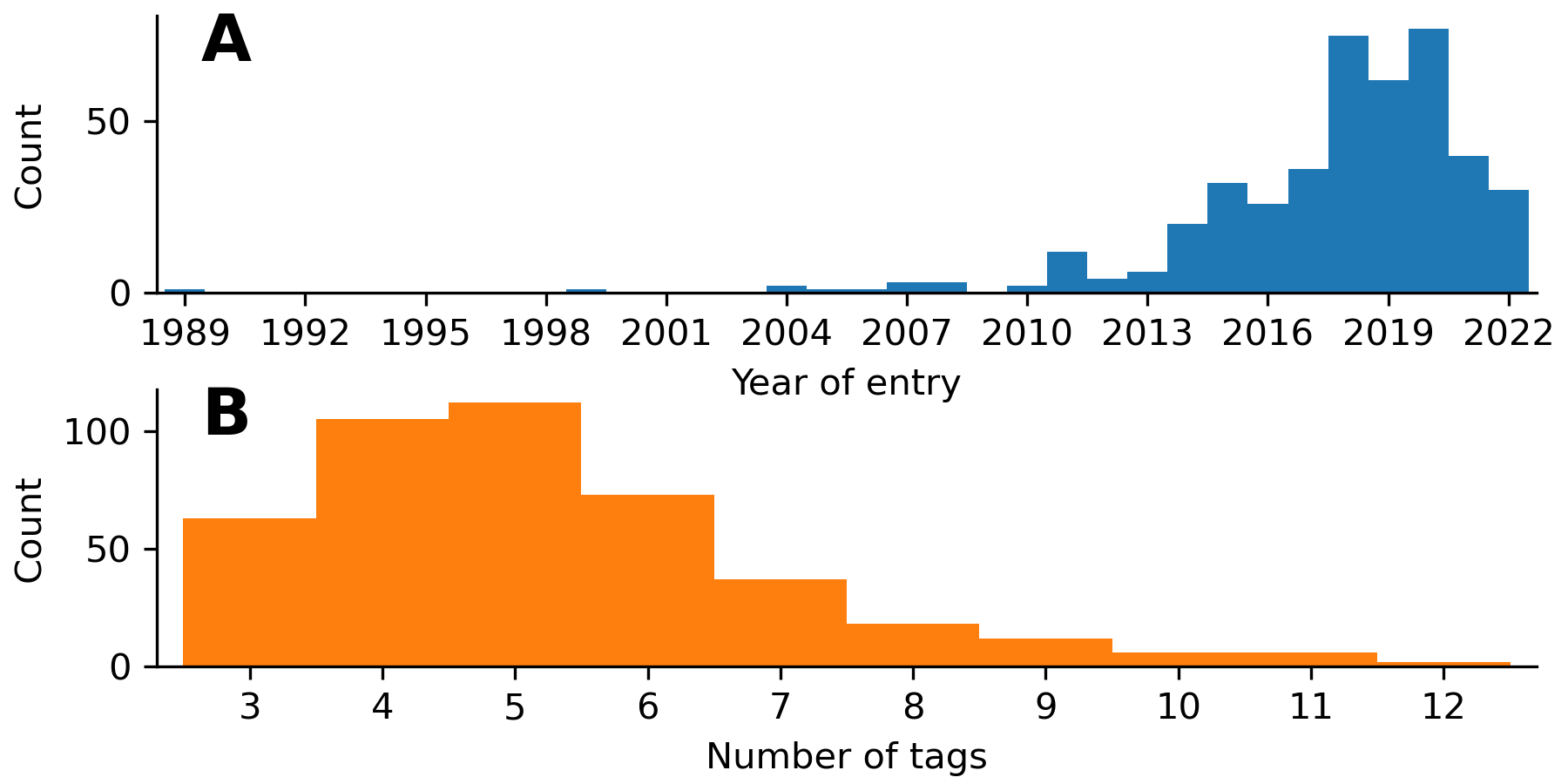}
\caption{Histograms of (A) the year of publication and (B) the number of tags for each entry}
\label{fig:histograms}
\end{figure}

\subsection{Tags \& Flavors}\label{s_tags}

There are three classes of tags (\textit{area of applications}, \textit{mathematical tools used}, and \textit{input type}) and two flavor labels (\textbf{innovate} and \textbf{confirm}). 
Every indexed entry must have at least one tag for each class; this is a hard requirement to ensure the utility of DONUT. 

\paragraph{\textit{Area of applications}.}
This is the most difficult tag to add. We hope to harness feedback from the community to continuously refine rules on adding this tag. Ideally, this tag should involve different scales, to ensure optimal search reults. 
For example, an entry about epilepsy should have as area-of-applications tag \texttt{medicine}, as the general field, \texttt{neurology} as the specification of it, and, finally, the most precise tag \texttt{epilepsy}. 
Because of how DONUT works (see~\Cref{S_technical}), searching for ``\texttt{epilepsy}'' and not for ``\texttt{tag:epilepsy}'' will result in all entries that mention the word, even if only in the bibliography, and thus in an imprecise search output.

\paragraph{\textit{Mathematical tools used}.} 
This class is easiest to tag, as authors are usually clear about the technical description of the analysis process and state the used tools explicitly. 
We aim to tag all employed tools, not just the ones from TDA, to provide a faithful summary of the context in which TDA is applied. 

\begin{figure}[h!]
\includegraphics[width=\columnwidth]{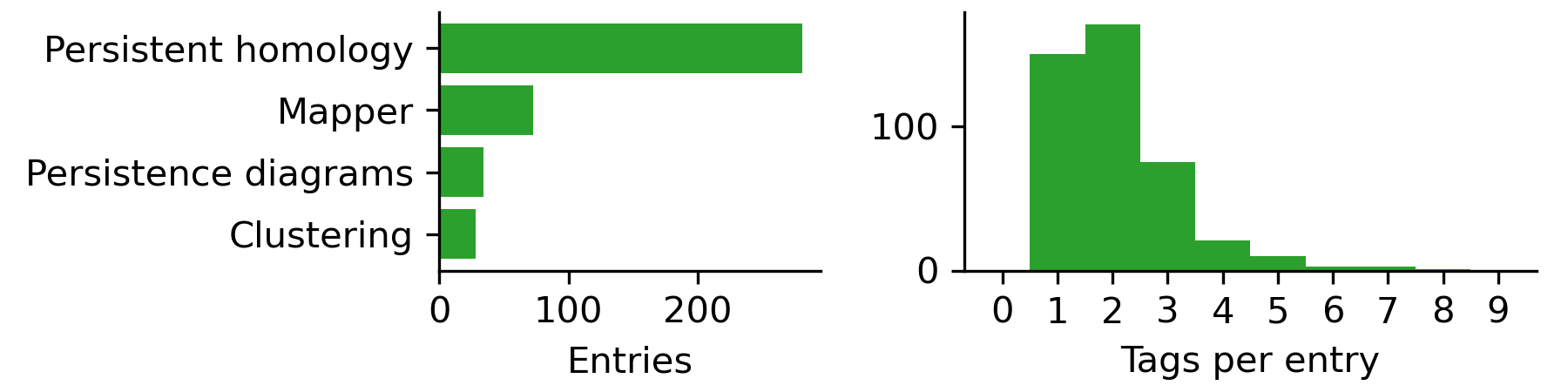}
\caption{The most popular tags (left) for the type of tool used, and a histogram (right) of how many tags of this type each entry has.}
\label{fig:popular_tags}
\end{figure}

\paragraph{\textit{Input type}.}
The third class of tag is conceptually very simple and denotes the data type(s) used in the TDA pipeline, such as gray-scale image, point cloud, time series, ... . 
However, in practice, we find that this tag is difficult to apply as input data usage is often not explicitly stated by the authors, or stated indirectly.
Analysis should always be reproducible, and authors failing to provide how the raw data were pre-processed to become suitable TDA input severely hinders reproducibility.

\paragraph{Flavors.} 
Flavors are not mandatory, both because their classification is more delicate and because not all entries fall clearly in one or the other type. 
The label \textbf{confirm} states that the findings of the entry are aligned with the outputs of existing methods. 
These entries perform the crucial job of reproducing results, and show that TDA can be used as an alternative, possibly better, method. 
The label \textbf{innovate} encodes all entries whose results are novel to the specific area of application. 
A result can be novel because obtained from larger datasets (too big to be handled by other methods), or because TDA extracts more information from the same data, or also because TDA can analyze data that other methods could not. 
Since this tag is critical in promoting TDA, we decided to be strict about it: if a result is not unquestionably novel, the flavor is \textit{not} added (58 entries out of the 431 cataloged are labeled \textbf{innovate}).

\section{Technical Details}\label{S_technical}

DONUT is based on a database of bibliography entries that is maintained via Zotero. The advantages of using Zotero are
\begin{inparaenum}[(i)]
  \item it provides a simple way of searching for publications and indexing them, and
  \item all bibliographical entries are stored as \hologo{BibTeX} entries.
\end{inparaenum}
This means that DONUT remains flexible and can be easily switched to another data source in the future, while at the same time not having to worry about issues with data entry.
Thus, DONUT consists of three independent components:
\begin{compactenum}
  \item An \emph{importer} for one-way synchronization between Zotero and the database of entries.
  \item A fulltext search engine for handling queries and maintaining the entries.
  \item A web frontend for interaction with the fulltext search engine.
\end{compactenum}
The importer is realized as a stand-alone program, making use of the Zotero API via \texttt{Pyzotero}~\cite{Pyzotero}.
The result of the parse process is a sequence of \hologo{BibTeX} entries.
Each of these entries are then inserted into \texttt{Xapian}, an open source full-text search engine.
\texttt{Xapian} indexes bibliographic information of documents and makes them accessible via well-defined API.
Finally, a web interface based on \texttt{Flask}, a Python frontend for web development, interacts with the database, depicts the results, and renders all queries.
Subsequently, we will briefly comment on the choice behind the search engine and the frontend.

\subsection{A Full-Text Search Engine}
The benefit of a full-text search engine like \texttt{Xapian} is that the indexing process of structured document data is full of hidden complexities.
For instance, are ``high-dimensional'' and ``high dimensional'' the same word?
How are simple spelling mistakes such as ``simplical'' instead of ``simplicial'' handled?
How are transliterated spellings~(``Pawel'' instead of Paweł) or approximate spellings~(``Pavel'') treated?
The frontend by Zotero ignores such questions and only permits simple queries that match a given string perfectly. \texttt{Xapian}, by contrast, is language-aware and can be set up to permit alternative spelling suggestions for queries.
Since the utility of DONUT hinges on the quality of its results for a given query string, we opted to index as much information about a bibliographic entry as possible.
As a result, DONUT is able to find documents more quickly than Zotero~(with query times ranging in the lower millisecond range) and provide more depth to queries.

\subsection{Frontend}
Users interact with databases typically through specialized query interfaces that are, ideally, as easy to use as Google.
Using \texttt{Flask}, a Python-based web framework, we provide such an interface~(in some sense, end users might perceive DONUT to \emph{be} the web interface, but as outlined above, DONUT actually consists of multiple parts).
The design choices behind the interface are first and foremost driven by speed and simplicity, following a minimalist design philosophy.
The search interface will work well on big screens and small screens alike, and care has been taken to follow official accessibility guidelines.
%
\begin{figure}[h!]
\includegraphics[width=\columnwidth]{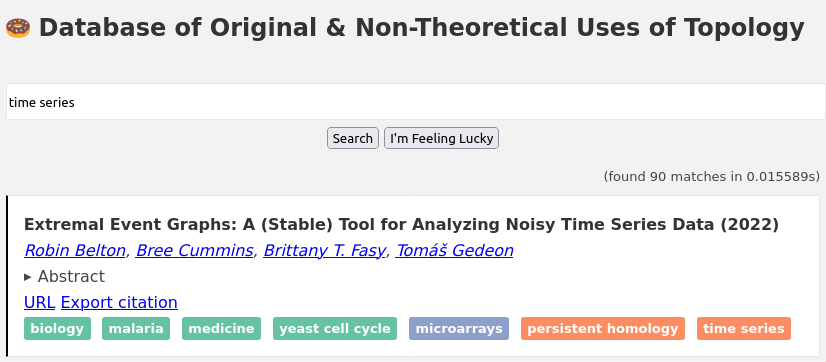}
\caption{User interface with a search term (only top result shown).}
\label{fig:screenshot}
\end{figure}

DONUT does not track users by means of cookies or related technologies.
General web server logs are stored in anonymized form, making it impossible to identify users.
These logs are used for diagnosing problems and providing summary statistics about accesses to the database.
Logs are stored encrypted and are automatically deleted on a rolling basis.
DONUT is thus fully compliant with GDPR and goes well beyond the ``best practices'' of contemporary websites.

\subsection{Code}

We make the code for DONUT available using GitHub, using a BSD 3-Clause License\footnote{
  \url{https://github.com/Pseudomanifold/DONUT}
}.
This license essentially permits anyone to use the code and modify it, provided our original copyright notice remains intact.
By making the code publicly available, any missing functionalities of the frontend or backend can be raised easily and addressed by the community or us.
We also hope that DONUT will inspire similar initiatives, since the code is not specific to applications in topology and could easily accommodate other scientific domains as well.

\subsection{Example Queries}

The frontend of DONUT can be used just like a regular search engine would be used.
Any queries are searched for in any of the fields of the database. Thus, searching for \verb|general| will find an entry called ``The Euler Characteristic: A General Topological Descriptor for Complex Data,'' but also all documents that have \verb|general| somewhere in their abstract, for instance.
This mimics the default behavior of search engines, which do not care about the location of a search term within a website.
To accommodate the needs of researchers, the DONUT frontend supports more refined queries and various operators.
For instance, searching for 
\verb|title:"general"| will only return the entry  ``The Euler Characteristic: A General Topological Descriptor for Complex Data,''
since~(at the time of writing this) no other entry has ``general'' in its title.
Similarly, one can search for document tags and authors, via \verb|tag:| and \verb|author:| respectively.

\paragraph{\textit{Automated normalization of queries}.}
Concerning the aforementioned issues of different spellings, one prominent feature of 
the query interface is that it supports transliterated spellings of names out of the box.
Hence, to stay with the original example, the search term \verb|author:pawel| will show results that include both the spelling ``Paweł,'' as well as the spelling ``Pawel.''
Similar rules apply to names with umlauts and other special characters.
This functionality is \emph{unique} to DONUT and not provided by the Zotero query interface.

\paragraph{\textit{Spelling suggestions}.}
Moreover, DONUT is capable~(in contrast to Zotero) to suggest other search terms to users based on similarity. Notice that DONUT will \emph{never} change the search query on its own. It will, however, suggest alternative concepts or spellings. For instance, searching for \verb|homotopy| brings up \verb|homology| as a potentially related query. Searching for \verb|homollogy|, on the other hand, will bring up no results, prompting DONUT to suggest ``Did you mean `\verb|homology|'?''.
We expect to further improve this functionality over time.

\subsection{Experimental Features}

We also use DONUT as a platform to experiment with various ways of making application papers more accessible and queryable.
For instance, we provide a ``landscape visualization''
that shows all indexed documents, following a natural landscape metaphor~\cite{Fabrikant10a}.
This provides a way to interact with documents and potentially find similar papers.

Another ongoing improvement  involves the indexing of the text of open-access publications.
This is considerably more complicated than integrating overall bibliographic information~(authors, abstracts, \dots) since it requires being able to process PDF files.
For entries whose text we can successfully process, we will incorporate the text in the search engine, meaning that keywords or phrases that appear only in the text of the entry will be made accessible to readers.
To ensure compliance with copyright, we will only do this for open-access publications.

\section{Future Opportunities}

We want DONUT first and foremost to be a useful tool \emph{by} the community \emph{for} the community.
As such, we believe that the most useful opportunities consist in expanding the taxonomy, that is, expanding the tagging system.
Going from ``more general'' to ``more specific'' leads to a natural hierarchy of tags.
For instance, an entry whose data tag is \verb|graphs| could be assigned a more specific tag of the form \verb|graphs:directed| if that is its context.
Over time, we expect that such a hierarchy will become more refined, allowing both unspecific and highly specific queries.
To aid users in their interactions with the hierarchy, we plan on implementing a ``tree visualization'' of tags.
When viewing an individual entry, we will make excerpts of the hierarchy visible, making it possible for users to navigate within the tree.

We hope that DONUT continues to be a useful tool for our community. Everyone is warmly invited to contribute to DONUT in various ways. We are open to additional suggestions for inclusion, updates to the web interface, as well as suggestions for new functionality.

\paragraph{\textit{Acknowledgments}.} The authors would like to thank the Hausdorff Research Institute for Mathematics for organizing excellent mathematical events, including the one at which the idea for this database was born, and Prof.\ Mikael Vejdemo-Johansson and Prof.\ Nina Otter for their contributions to the database.
B.R.\ is grateful for discussions with Lukas Hahn, 
Maximilian Schmahl, and Daniel Spitz.
B.G.\ was supported by the Austrian Science Fund (FWF) P 33765-N. 

\bibliography{Refs}

\end{document}